\documentclass[twocolumngrid,prd,superscriptsize,reprint]{revtex4-1}
\usepackage[]{graphicx}
\usepackage{dcolumn}
\usepackage{bm}

\usepackage{amsmath,amssymb,amsfonts}
\usepackage{fancyhdr}
\usepackage{lipsum} 
\usepackage{subcaption}
\usepackage{multirow}

 \usepackage[colorlinks]{hyperref}
 \hypersetup{
     colorlinks = false,
   }

\newcommand{\comment}[1]{}

\usepackage{color}
\usepackage{babel}
\newcommand{\bea}{\begin{eqnarray}}
\newcommand{\eea}{\end{eqnarray}}

\newcommand{\be}{\begin{equation}}
\newcommand{\ee}{\end{equation}}

\begin{document}

\title[]{Relativistic Moduli Space and critical velocity in kink collisions}
\author{C. Adam}
\affiliation{Departamento de F\'isica de Part\'iculas, Universidad de
Santiago de Compostela and \\
Instituto Galego de F\'isica de Altas Enerxias (IGFAE), E-15782
Santiago de Compostela, Spain}
\author{D. Ciurla}
\author{K. Oles}
\author{T. Romanczukiewicz}
\author{A. Wereszczynski}
\affiliation{Institute of Theoretical Physics, Jagiellonian University,
Lojasiewicza 11, Krak\'{o}w, Poland}

\begin{abstract}
We analyze the perturbative Relativistic Moduli Space approach, where the amplitudes of the Derrick modes are promoted to collective coordinates. In particular, we analyse the possibility to calculate the 
critical velocity, i.e., the initial velocity of kinks at which single bounce scattering changes into a multi-bounce or annihilation collision,
in the resulting Collective Coordinate Model (CCM). We find that for a growing number of modes the critical velocity of the CCM approaches the full field theory value. This is verified in the case of the $\phi^4$ model, where we reach a $99\%$ accuracy. We also see such a convergence for a wide range of models belonging to the family of the double sine-Gordon and Christ-Lee theories, especially in those cases where the kinks do not reveal a too well pronounced half-kink inner structure. 
\end{abstract}
\maketitle


\section{Introduction}


When topological kinks \cite{SM, Shnir} are collided, they show a very complicated behavior. They can (quasi) elastically scatter and reappear in the final state with smaller or bigger energy loss. They can also annihilate, that is, completely decay into small waves, i.e., radiation, often via the formation of an intermediate long-living quasi-periodic state called oscillon (or bion). These two scenarios are supplemented by the appearance of few-bounces, where the colliding kink and antikink meet a few times before they manage to escape to infinity \cite{sug, CSW, GH}. The resulting behavior in a particular process strongly depends on initial parameters like the velocities of the kinks \cite{sug, CSW, GH} or the amplitudes of additional excitations \cite{A}. This complex pattern of possible outcomes is encapsulated in the famous chaotic or even fractal structure in the final state formation \cite{sug, CSW}. 

The explanation of this involved, chaotic pattern even for the simplest, prototypical kink-antikink collision in $\phi^4$ theory, was a long standing challenge which only very recently led to a satisfactory resolution. Indeed, for more than 40 years it was expected than the fractal structure is related to the {\it resonant transfer mechanism} which is nothing but a flow of energy between kinetic and internal degrees of freedom (DoF) \cite{sug, CSW}. The kinetic DoF is just kinetic motion of the soliton connected with a zero mode arising from the translational invariance of the theory, while the internal DoF are typically normal or quasi-normal modes hosted by the solitons. In the simplest case, initially the colliding solitons have only kinetic energy which during the first collision is not only lost via the emission of radiation, but also excites the internal modes. This may result in a situation where the kink and antikink do not have enough kinetic motion to overcome their mutual attractive force and, therefore, have to collide again. Often, more collision means more radiation and further energy loss, but sometimes the energy stored in the internal DoF can return to the kinetic DoF, allowing the solitons to liberate and escape to infinities. 

This simple and beautiful mechanism explains, e.g., the frequency of the oscillations in two and even three bounce windows in the $\phi^4$ model. However, serious problems in the construction of a Collective Coordinate Model (CCM) based on these DoF \cite{Weigel}, cast some doubts on the resonant transfer mechanism and/or the CCM approach itself \cite{Kev}. Fortunately, it was recently shown that there exists a two-dimensional CCM based on only two collective coordinates, the position of the solitons and the amplitude of their internal mode (in this case called shape mode), which {\it qualitatively} reproduces the full field theory dynamics \cite{MORW}. 

This break through led to another important question, whether there may exist a CCM type approach which not only qualitatively agrees with the full theory, but which could be treated as a precision tool, allowing for a more detailed, {\it quantitative} agreement. In particular, whether there is a CCM scheme in which (at least some) observables converge to the values computed in the full field theory. 

The first step in this attempt has already been made, by the discovery of the so-called perturbative Relativistic Moduli Space (pRMS) approach, where instead of a very restricted number of normal bound modes one uses an, in principle, {\it arbitrary} number of {\it Derrick modes} \cite{RMS}. It has been shown that using two Derrick modes significantly improves the results of the CCM, indicating that such a converging framework may indeed exist. 

It is the main objective of the current work to further explore this possibility. In particular, we will show that one of the most important observables, i.e., the {\it critical velocity} which separates the region of the simplest one bounce scattering from multi-bounce/annihilation processes, can be very accurately captured by a CCM based on the pRMS, where the agreement gets better when we increase the number of Derrick modes. This will be established for a wide set of models, especially those for which kinks are localized around one center rather than two or more.  Thus, we find a first converging CCM like framework, at least for the critical velocity. 

\section{Perturbative Relativistic Moduli Space}
Let us begin with a general real scalar field theory in (1+1) dimensions 
\be
L=\int_{-\infty}^{\infty} \left( \frac{1}{2}\left( \partial_\mu \phi \right)^2 - U(\phi) \right)dx,
\ee
where the field theoretical potential $U$ defines a particular theory and has at least two vacua. This allows for the existence of topologically nontrivial solitons called (anti)kinks $\Phi_{K(\bar{K})}(x;a)$ which obey the well-known static Bogomolny equation 
\be
\frac{d \phi}{dx} = \pm \sqrt{U(\phi)}.
\ee
Here $a \in \mathbb{R}$ is the position of the soliton and its arbitrariness reflects the translational invariance of the field theory. 
Obviously, a static (anti)kink can be boosted in a Lorentz covariant way. Unfortunately, there are typically no analytical solutions describing a dynamical process involving both kink and antikink and, apart from the numerical treatment of the full equations of motion, not many methods are available. One important exception is the collective coordinate model (CCM) method which allows to reduce the infinitely many DoF of the original field theory to a discrete set of parameters, i.e., {\it moduli}, whose time evolution approximates the full PDE dynamics, see e.g.,  \cite{NM-2, NM-1}. 

Concretely, the infinite dimensional space of field configurations is reduced to an $N$-dimensional subspace spanned by a restricted set of configurations $\mathcal{M}(X^i)=\{ \Phi(x; X^i), i=1..N\}$. The identification of such a set is usually a very nontrivial task. In the next step, the continuous parameters $X^i$ are promoted to time dependent coordinates providing a mechanical-like system 
\be
L[{\bf X}]=\int_{-\infty}^\infty  \mathcal{L}[\Phi(x; X^i(t))] \, dx
= \frac{1}{2} g_{ij}({\bf X}) \dot{X}^i \dot{X}^j - V({\bf X}) \,,
\label{eff-lag}
\ee
where
\be
g_{ij}({\bf X})=\int_{-\infty}^\infty \frac{\partial \Phi}
{\partial X^i} \frac{\partial \Phi}{\partial X^j} \, dx
\label{modmetric}
\ee
is the metric on the moduli space $\mathcal{M}$ and
\be
V({\bf X})=\int_{-\infty}^\infty \left( \frac{1}{2}
\left( \frac{\partial \Phi}{\partial x}
\right)^2 + U(\Phi) \right) \, dx
\label{modpot}
\ee
is an effective potential. 

This approach works especially well for BPS models where there is no static force between solitons. E.g., it explains the famous $\pi/2$ scattering of the Abelian Higgs vortices at critical coupling, or of BPS monopoles \cite{NM-2, NM-1}. Moreover, it is also very useful in non-BPS theories, where it gives a strong evidence that certain effects arising in soliton dynamics result from a coupling between kinetic and internal DoF. See, for example, the fractal structure in the final state formation in kink-antikink collisions in the $\phi^4$ model \cite{MORW}. Finally, the identification of the correct collective coordinates is of high importance for the semiclassical quantization of solitons \cite{Sam, RL, BS, CH}.

Typically, the construction of a moduli space for multi-kink processes begins with single kink states, $\Phi_{K(\bar{K})}(x; a)$, which are then trivially superposed. For a realistic description of non-BPS processes, the inclusion only of the single (anti)kink solutions at different positions $a$ is not enough and one needs to add single soliton modes $\eta^{K(\bar{K})}_i(x; a)$. For example, for symmetric kink-antikink (KAK) collisions we choose  
\bea
\Phi_{K\bar{K}}(x;a,{\bf X}) &=& \Phi_{K}(x; -a) + \Phi_{\bar{K}}(x; a) + \Phi_{vac}\nonumber \\ 
&+& \sum_{i=1}^N X^i \left( \eta^{K}_i(x; -a) + \eta^{\bar{K}}_i(x; a) \right),
\eea
where $\Phi_{vac}$ is a vacuum value added to provide the correct boundary conditions. 
Sometimes, even multi-soliton (delocalized) modes have to be included \cite{phi6}. 

It is important to notice that in this construction the original Lorentz covariant field theory is lost, and we arrive at a non-relativistic CCM \cite{SM}. This is a valid approximation if the velocity of the solitons is small, otherwise a more refined approach is needed.

Such a construction exists and requires taking into account at least one additional, scale (Derrick) deformation, $x\to bx$, where the scale factor $b \in \mathbb{R}_+$ \cite{Rise, Caputo}. In the single soliton sector it gives the following moduli space
\be
\mathcal{M}_{K(\bar{K})}(a,b)= \{ \Phi_{K(\bar{K})} (b(x-a))\}. 
\ee
The resulting CCM possesses a stationary solution 
\be
\dot{a}=v, \;\;\; b=\frac{1}{\sqrt{1-v^2}},
\ee
which describes the Lorentz contraction of a boosted (anti)kink. 

This nice relativistic generalization meets, unfortunately, serious difficulties if applied to KAK collisions in theories where the kink and antikink are related by a simple change of sign  
\be
\Phi_{\bar{K}}(x)=-\Phi_K(x),
\ee
which happens, for example, in the $\phi^4$ and sine-Gordon models. Indeed, then the KAK moduli space 
\be
\mathcal{M}_{K\bar{K}}(a,b)= \{ \Phi_{K} (b(x+a)) - \Phi_{K} (b(x-a))\}
\ee
reveals a singularity at $a=0$, where two cone like surfaces (for $a>0$ and $a<0$) are joined. This is a genuine singularity which cannot be removed by any change of coordinates. As a consequence, the resulting CCM collapses at this point \cite{RMS}. 

To cure this obstacle, a perturbative scheme called the {\it perturbative Relativistic Moduli Space} approach (pRMS) has been introduced recently \cite{RMS}. As its name suggests, the Derrick deformation is treated perturbatively, i.e., $b=1+\epsilon$, where $\epsilon$ is formally a small parameter. Then, we insert it into the single kink solution and expand it in powers of $\epsilon$ to an arbitrary order $N$
\be
\Phi_K(b(x-a))=\sum_{k=0}^N \frac{\epsilon^k}{k!} (x-a)^k \Phi_K^{(k)} (x-a) + o(\epsilon^N),
\ee
where $\Phi^{(k)}$ is the $k$th derivative of the kink. Then we do the crucial step and treat each term in the expansion as an independent deformation, the $k$th Derrick mode. This means that we replace $\epsilon^k$ by a new, independent collective coordinate $C_k$ obtaining the following restricted set of configurations
\be
\Phi_K(x; a, {\bf C})= \Phi_K(x-a)+\sum_{k=1}^N \frac{C_k}{k!} (x-a)^k \Phi_K^{(k)} (x-a) \label{1-kink-pert}.
\ee
Importantly, this framework introduces an {\it arbitrary} number of collective coordinates. The first few may be assigned to normal modes (shape modes) of the kink, while higher ones have frequencies which are above the mass threshold and, therefore, can be associated with a sort of radiation, at least at short time scales. Of course, they do not represent true radiation, since all Derrick modes are bounded to the kink and cannot escape to infinite. 

Similarly as in the non-perturbative relativistic set-up, there is a solution of the single soliton sector which describes an approximation of a boosted kink. Indeed, a CCM based on (\ref{1-kink-pert}) has a stationary solution 
\be
\dot{a}=v, \;\;\; C^k=\tilde{C}^k, \label{stationary}
\ee
where the $\tilde{C}^k$ solve the following algebraic equations 
\be
\frac{v^2}{2} \frac{\partial g_{aa}({\bf C})}{\partial C^k} = \frac{\partial V({\bf C})}{\partial C^k}.
\ee

It is straightforward to construct the KAK moduli space. Namely, it reads
\bea
\Phi_{K\bar{K}}(x; a, {\bf C}) &=& \Phi_K(x-a) - \Phi_K(x-a) +\Phi_{vac} \nonumber \\
&+&\sum_{k=1}^N \frac{C_k}{k!} \left( (x+a)^k \Phi_K^{(k)} (x+a) \right. \nonumber \\
&-& \left. (x-a)^k \Phi_K^{(k)} (x-a) \right). \label{pRMS-set}
\eea
It, again, leads to a singularity on the moduli space at $a=0$, but now this is an apparent singularity which can be removed by a suitable change of coordinates \cite{MORW-2}. Concretely, we apply the following redefinition $C_k \to C_k / \tanh(a)$ and arrive at the following expression for the restricted set of configurations, 
\bea
\Phi_{K\bar{K}}(x; a, {\bf C}) &=& \Phi_K(x-a) - \Phi_K(x-a) +\Phi_{vac} \nonumber \\
&+&\sum_{k=1}^N \frac{C_k}{k! \tanh(a)} \left( (x+a)^k \Phi_K^{(k)} (x+a) \right.\nonumber \\
&-& \left. (x-a)^k \Phi_K^{(k)} (x-a) \right). 
\eea
which finally leads to a well defined CCM
\be
L[a,{\bf C}]=\int_{-\infty}^\infty \mathcal{L}[\Phi_{K\bar{K}}(x; a, {\bf C})] dx.
\ee 
This must be further equipped with suitable initial conditions. In all cases, we scatter initially unexcited solitons which are boosted towards each other with an initial velocity $v_{in}$. Thus, we inherit the initial conditions from the stationary solution of the single soliton CCM
\be
a(0)=a_0, \;\;\; \dot{a}(0)=v_{in}, \;\;\; C^k(0)=\tilde{C}^k(v_{in}), \;\;\; \dot{C}(0)=0,
\ee
where $a_0$ is half of the initial separation of the kink and the antikink. 

In the next sections, we will test this construction for three types of theories, the $\phi^4$, Christ-Lee and double sine-Gordon models. 
\section{$\phi^4$ model}
In our first example, we continue the previous study of KAK collisions in the $\phi^4$ model
\be
U_{\phi^4}= \frac{1}{2} (1-\phi^2)^2.
\ee
This is the prototypical kink process in a non-integrable theory whose explanation has been a challenge for many years. 

The kink and antikink at the origin are given as
\be
\Phi_{K(\bar{K})}(x)=\pm \tanh(x)
\ee
and they host only one bound mode, called shape mode, known in an exact form,
\be
\eta_{sh}(x)= \sqrt{\frac{3}{2}} \frac{\sinh(x)}{\cosh^2(x)} ,
\ee
with frequency $\omega_{sh}=\sqrt{3}$. At $\omega=4$ the continuum spectrum of scattering modes begins. 
\begin{figure*}
 \includegraphics[width=1.70\columnwidth]{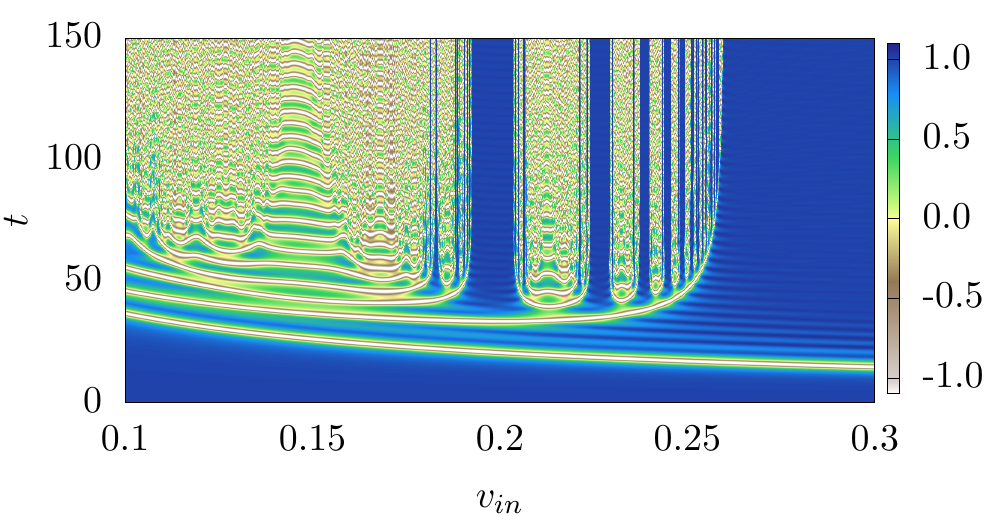} 
  \includegraphics[width=1.0\columnwidth]{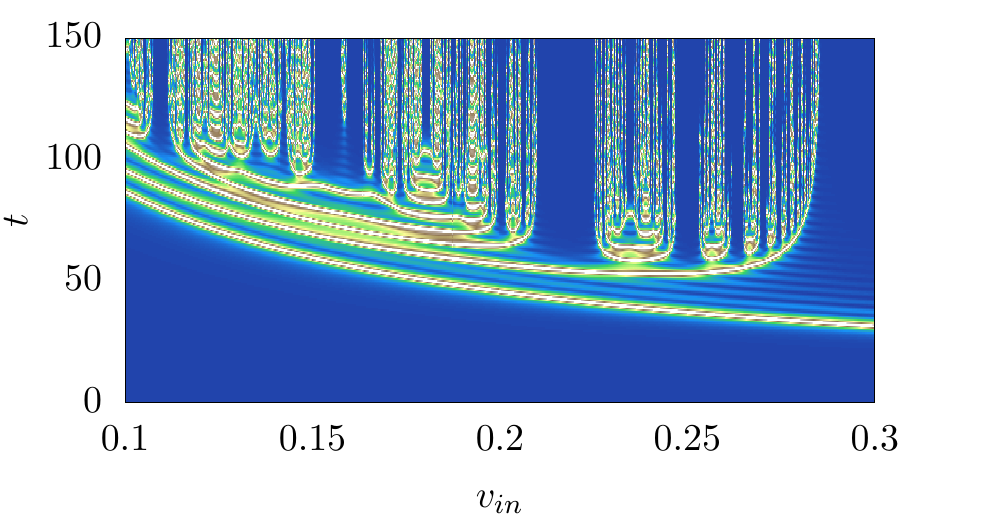} 
    \includegraphics[width=1.0\columnwidth]{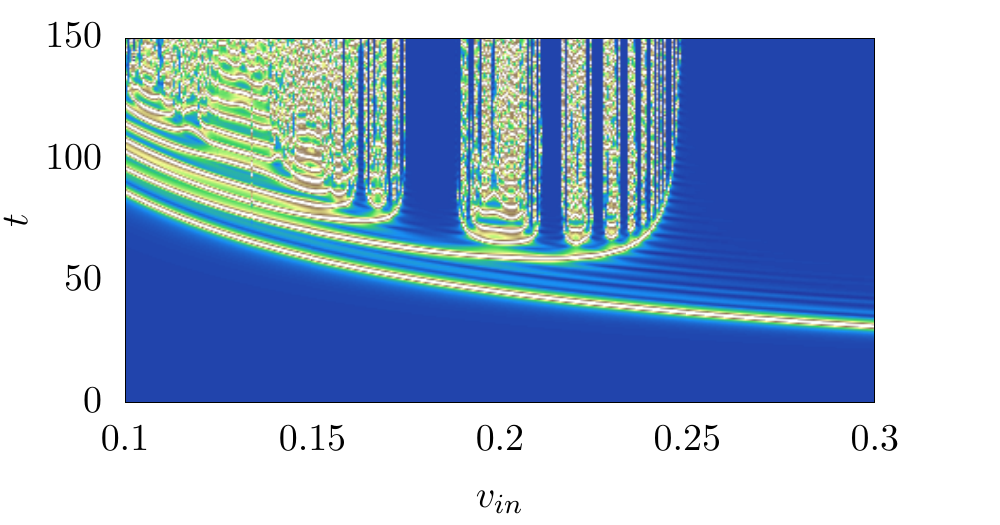} 
      \includegraphics[width=1.0\columnwidth]{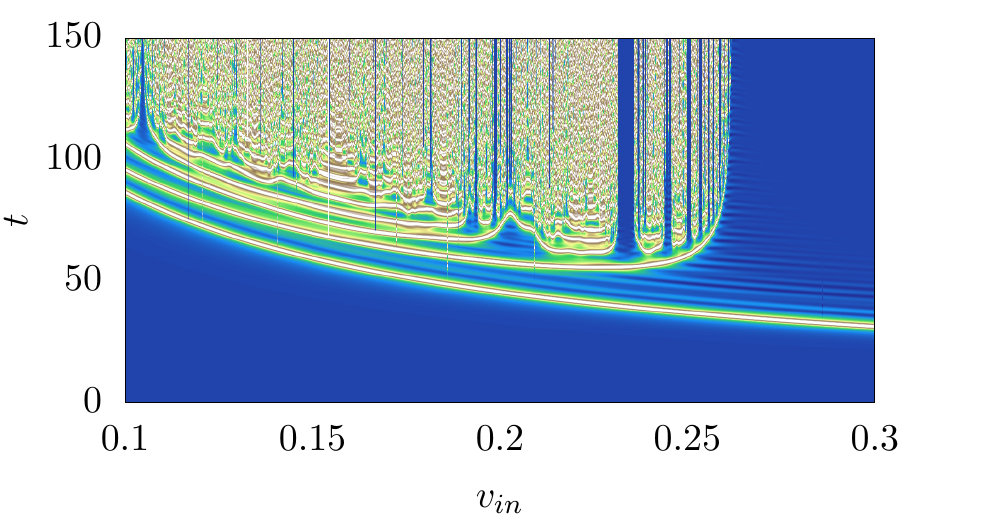} 
        \includegraphics[width=1.0\columnwidth]{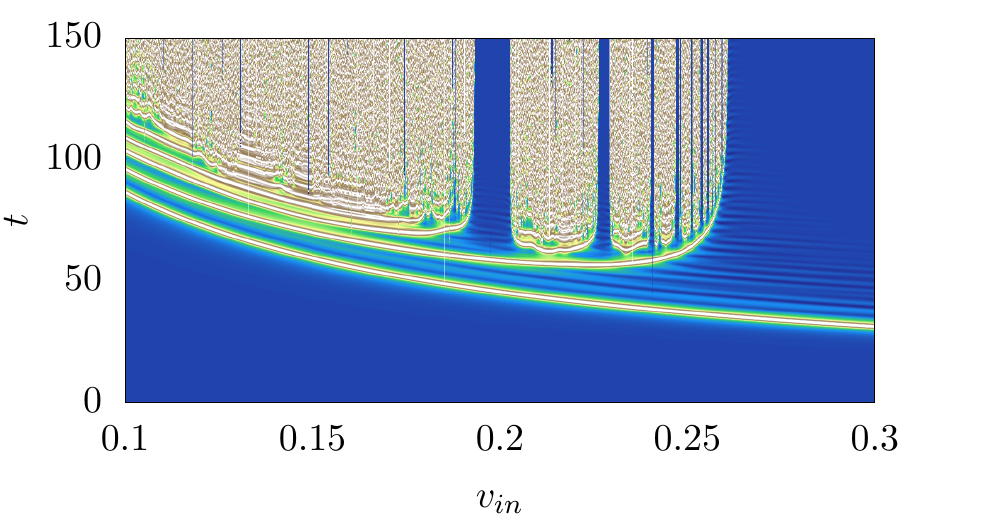} 
   \caption{Kink-antikink collision in the $\phi^4$ model: full theory (upper) and in CCM based on pRMS with one (central left), two (central right), three (lower left) and four (lower right) Derrick modes. } \label{KAK-plot}
 \end{figure*}

As is well known, the KAK scattering in the $\phi^4$ model exhibits a chaotic structure in the final state formation \cite{sug, CSW, GH, Kev}, where between the single-bounce scattering and the region of complete annihilation, one finds a very complicated, fractal pattern of few-bounce windows surrounded by other annihilation regions, see Fig. \ref{KAK-plot}, upper panel. In bounce windows, the solitons for certain collisions gain a sufficient amount of kinetic energy to overcome the attractive kink-antikink force and escape to infinities as free final states. In the annihilation regions, called bion chimneys, they form an imperfect version of the famous breather, i.e.,  oscillon, which here slowly decays to the vacuum by the emission of radiation. One very important observable is the {\it critical velocity} which divides the one-bounce collisions from the chaotic regime. In this case, $v_{crit}=0.2598$. We will model this highly complicated dynamics using the pRMS construction. 

In Fig. \ref{KAK-plot} we show the KAK dynamics obtained in the CCM based on the pRMS (\ref{pRMS-set}). We vary the number of Derrick modes from one to four. The results for $L[a,C^1]$ and $L[a,C^1,C^2]$ were originally presented in \cite{RMS}. We see that even the simplest CCM which contains only the first Derrick mode qualitatively reproduces the full PDE dynamics. This qualitative agreement improves significantly if we include the second Derrick mode, see Fig. \ref{KAK-plot}. Indeed, the previously observed unwanted three and four-bounce windows, which existed for $v_{in} \in (0.11,0.2)$, disappear. Also an overall shift of the bounce structure to larger velocities is not present any longer. In fact, the results become in {\it quantitative} agreement with the full field theory. This is a solid evidence that the resonant energy transfer involving the kinetic and internal DoF is responsible for the fractal structure observed in the formation of the final state. 

\begin{table}  \label{tab-f}
{\scriptsize 
\begin{tabular}{ccccc} 
\hline\hline
\hspace*{0.5cm}$N$ \hspace*{0.5cm} & \hspace*{0.5cm} $v_{crit}$ \hspace*{0.5cm} &  \hspace*{0.5cm}  agreement  $[\%]$ \hspace*{0.5cm}  \\
\hline
1 & $\texttt{0.2854} $ & $\texttt{90.15} $   \\
2 & $\texttt{0.2491} $ & $\texttt{95.88} $     \\
3 & $\texttt{0.2639} $ & $\texttt{98.42} $  \\
4 & $\texttt{0.2618} $ & $\texttt{99.23} $  \\
\hline\hline
\end{tabular}} 
\caption{Comparison of the critical velocities for the KAK collision in $\phi^4$ model obtained in the CCM based on $N=1,2,3$ and $4$ Derrick modes.  The true critical velocity is $v_{cr}=0.2598$.}
\end{table} 

We also note that the second Derrick mode may to some extent play the role of radiation since its frequency lies above the mass threshold, $\omega^2_2=6.9283$. Also the frequency of the first Derrick mode is closer to the frequency of the shape mode if we include the second Derrick mode. Namely, it shrinks from $\omega^2_1=3.1011$ (only the first Derrick mode included) to $\omega^2_1=3.0221$.

 If we include the third Derrick mode, even more bounce windows disappear and the results seem to be a little bit worse than in the case with two Derrick modes, Fig. \ref{KAK-plot}, lower left panel. The picture improves in we consider four Derrick modes, Fig. \ref{KAK-plot} lower right panel.  The overall tendency is that higher rank bounce windows are suppressed and the CCM scattering is dominated by one, two or three bounce windows and bion chimneys. Thus, from the point of view of higher-bounce windows, the inclusion of too many Derrick modes does not seem to be an optimal strategy. 

There is, however, an amazing improvement in the prediction of the critical velocity if we increase the number of Derrick modes, see Tab. I. The agreement between the CCM models and the full field theory grows from $90\%$ for one Derrick mode to a striking $99\%$ if four Derrick modes are taken into account. 

\section{Time dynamics of Derrick modes}
In fact, the problem in the description of higher rank bounces as well as the significant improvement of the prediction of the critical velocity, may have the same origin. Namely, both seem to be related to the fact that the higher Derrick modes can be used to approximate radiation only in a short time scale. Initially, they effectively transfer energy from the kinetic motion and the first Derrick mode (which acts as the shape mode). However, in contrast to radiation, the higher Derrick modes are also confined to the (anti)kink. Thus, the fraction of energy stored in those modes is never released to infinity but, after not so long time, can enter again the center of the soliton, leading to an additional excitation of the zero and/or the first Derrick mode. Thus, medium and large time effects may be affected by this "bounded radiation" effectively provided by the higher order Derrick modes. 

\begin{figure}
 \includegraphics[width=1.0\columnwidth]{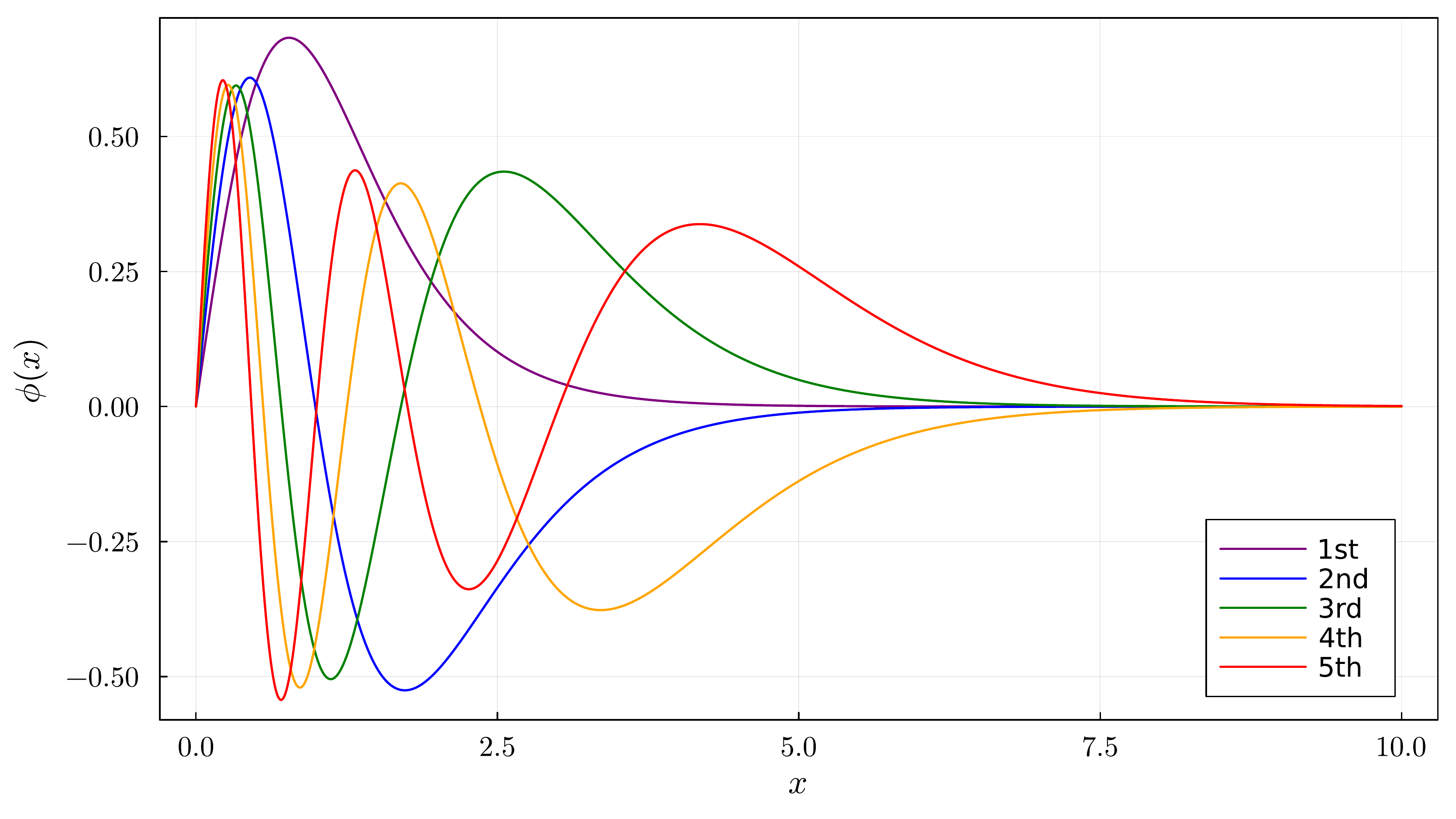} 
   \caption{The first five Derrick modes of the $\phi^4$ kink.} \label{4-derricks}
    \includegraphics[width=1.0\columnwidth]{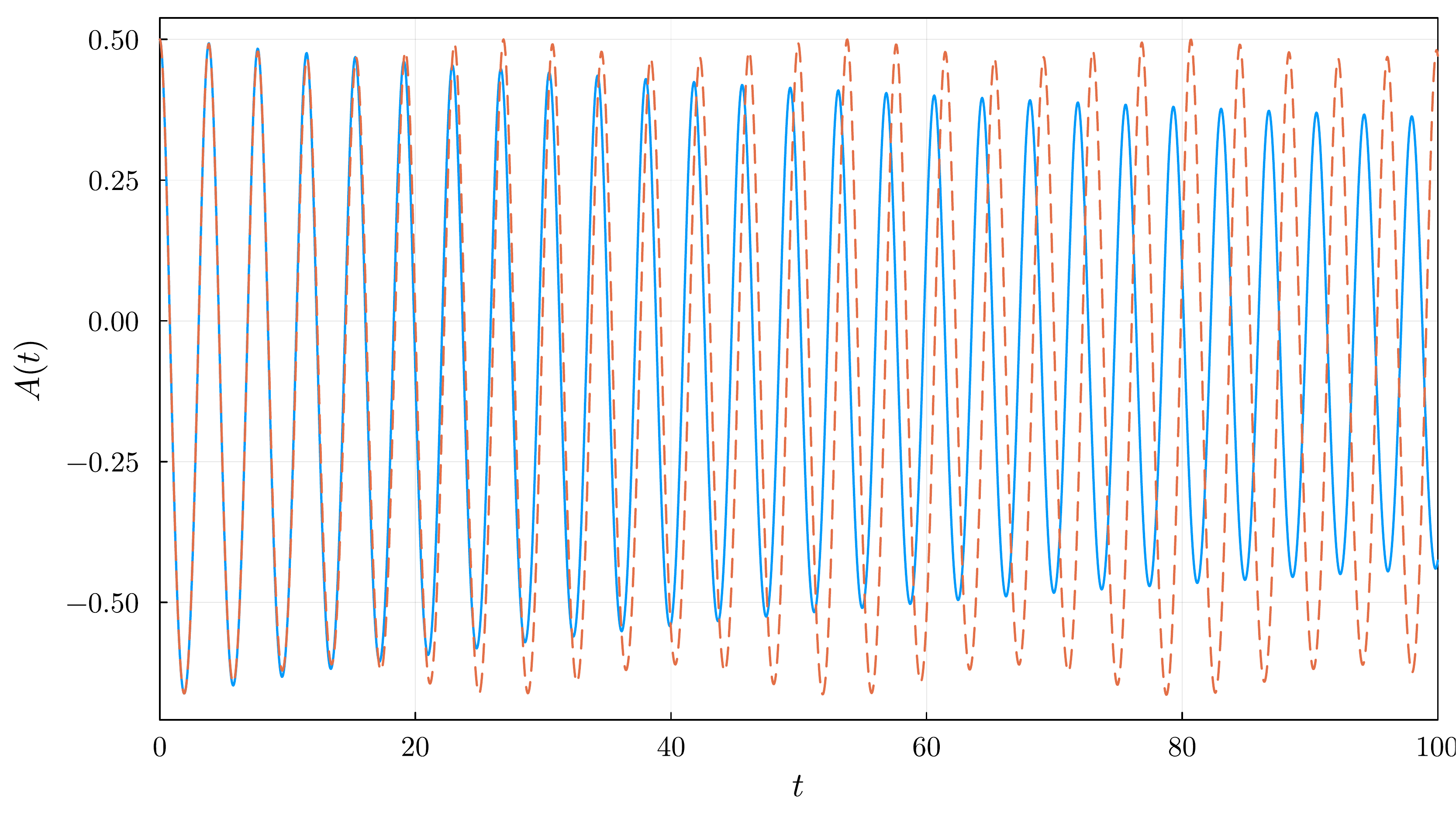}
   \caption{Shape mode evolution in the $\phi^4$ model (blue curve) and in the CCM based on five Derrick modes (orange).} \label{4-mode-evol}
  \includegraphics[width=1.0\columnwidth]{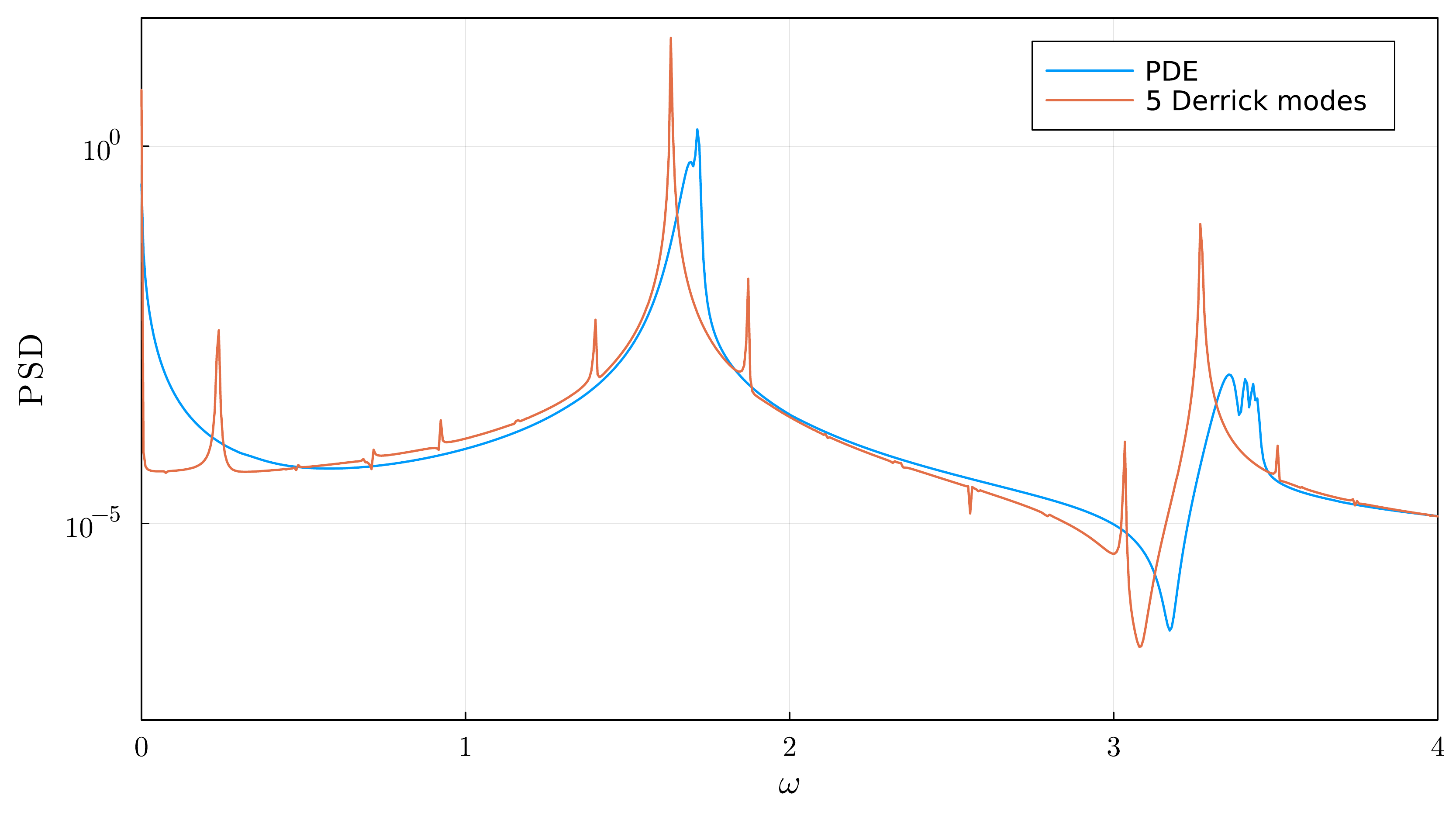}
   \caption{Power spectrum of the evolution of the shape mode.} \label{4-mode-power}
 \end{figure}

On the other hand, the critical velocity divides the regime with only one bounce from a more complicated behaviour. Thus, it is controlled by the behaviour of the kinks at the first bounce, that is, in the short time interval where the Derrick modes are still trustable. 

To verify this heuristic argumentation, we analyze the time dynamics of the Derrick modes in the single soliton sector. First of all, let us observe that higher rank Derrick modes are more spread. This is shown in Fig. \ref{4-derricks}. That was the reason why these modes can be treated as a surrogate of radiation. Namely, they can transfer energy from the center of the kink. 

However, as is shown in Fig. \ref{4-mode-evol}, such an energy transfer occurs only for not too long time scales. Here we show the decay of the shape mode in the full field theory, and the one obtained in the single kink CCM model with the first five Derrick modes included. The initial shape mode amplitude is $A(0)=0.5$. As proved in \cite{MM}, the shape mode decays due to nonlinearities of the $\phi^4$ model, which couples the normal mode to radiative modes. This leads to a $t^{-1/2}$ decay of the shape mode amplitude at large time. In the CCM approach, we initially see a decay of the shape mode amplitude due to the flow of the energy to higher Derrick modes. This happens for a few shape mode oscillations. For later time, we observe the reversed energy transfer, and the amplitude of the shape mode increases. This is then repeated many times, leading to an apparent double oscillation structure. This may result in a too high value of the shape mode amplitude at the second and higher bounces, which at the end can be a source of the growing disagreement in their description. 

The emergence of long time scale oscillations in the CCM approximation of the shape mode is also clearly visible in Fig. \ref{4-mode-power}, where we plot the power spectrum for the solutions of Fig. \ref{4-mode-evol}. Indeed, beside a very good agreement in the frequency of the shape mode (and its higher harmonics) we see a peak obtained for the CCM dynamics at a small frequency.  

\section{Christ-Lee Model}

Now, we turn to a family of theories known as the Christ-Lee model \cite{CL},
\be
U_{CL}=\frac{1}{2(1+\epsilon^2)} (\epsilon^2+\phi^2)(1-\phi^2)^2.
\ee
This is a version of a sixth order potential which was extensively analyzed as a (1+1) dimensional analog of the bag model, with kinks playing the role of confined quarks in a solitonic baryon. 
\begin{figure}
 \includegraphics[width=1.10\columnwidth]{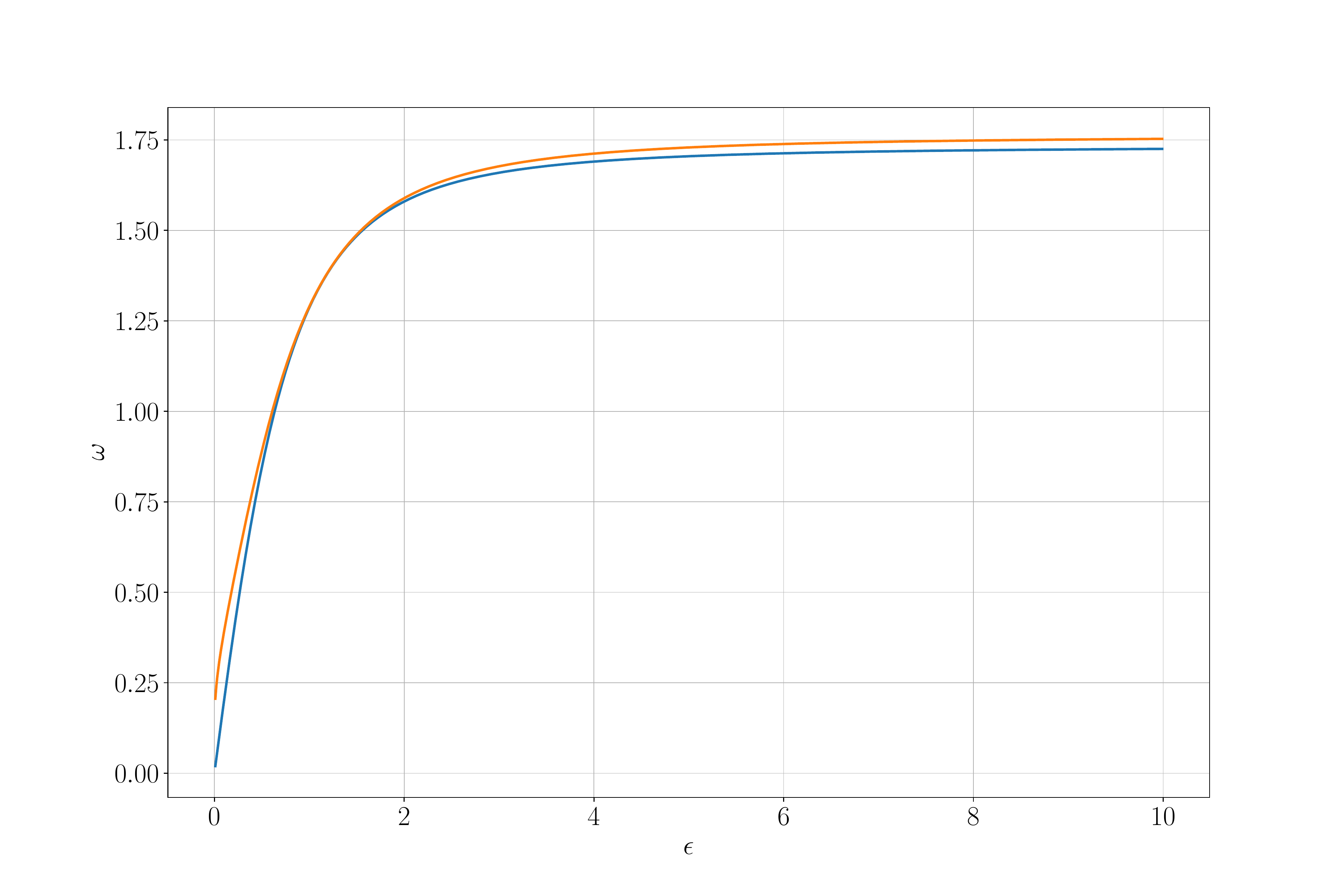} 
   \caption{The lowest frequency bound mode in the Christ-Lee model (blue) and the first Derrick mode (orange). } \label{CL-mode}
 \end{figure}
\begin{figure}
 \includegraphics[width=1.00\columnwidth]{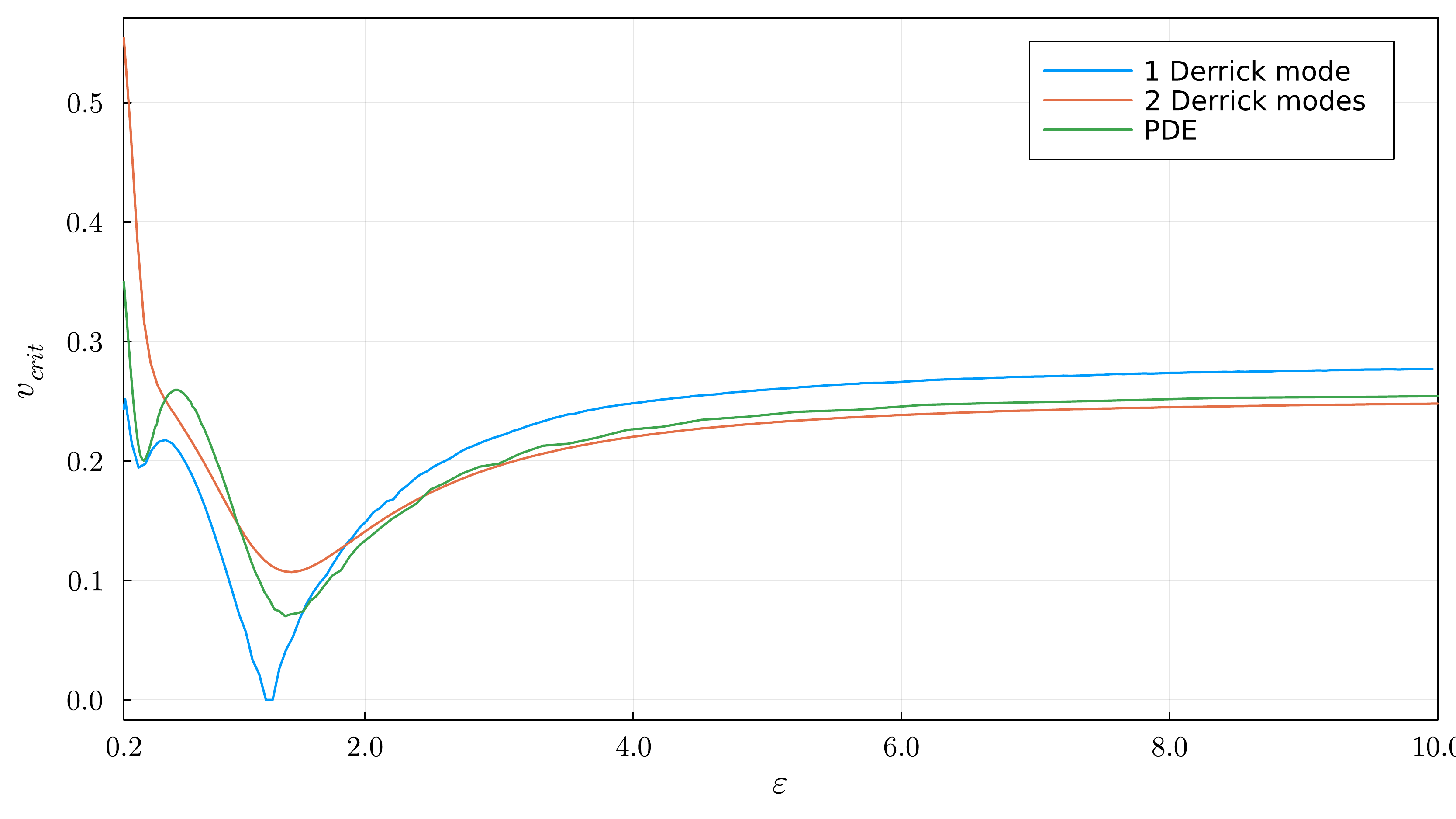} 
   \caption{The critical velocity $v_{cr}$ in the Christ-Lee model (green) and result obtained from the CCM based on one (blue) and two (orange) Derrick mode(s). } \label{CL-vcrit}
 \end{figure}

\begin{figure*}
  \includegraphics[width=2.0\columnwidth]{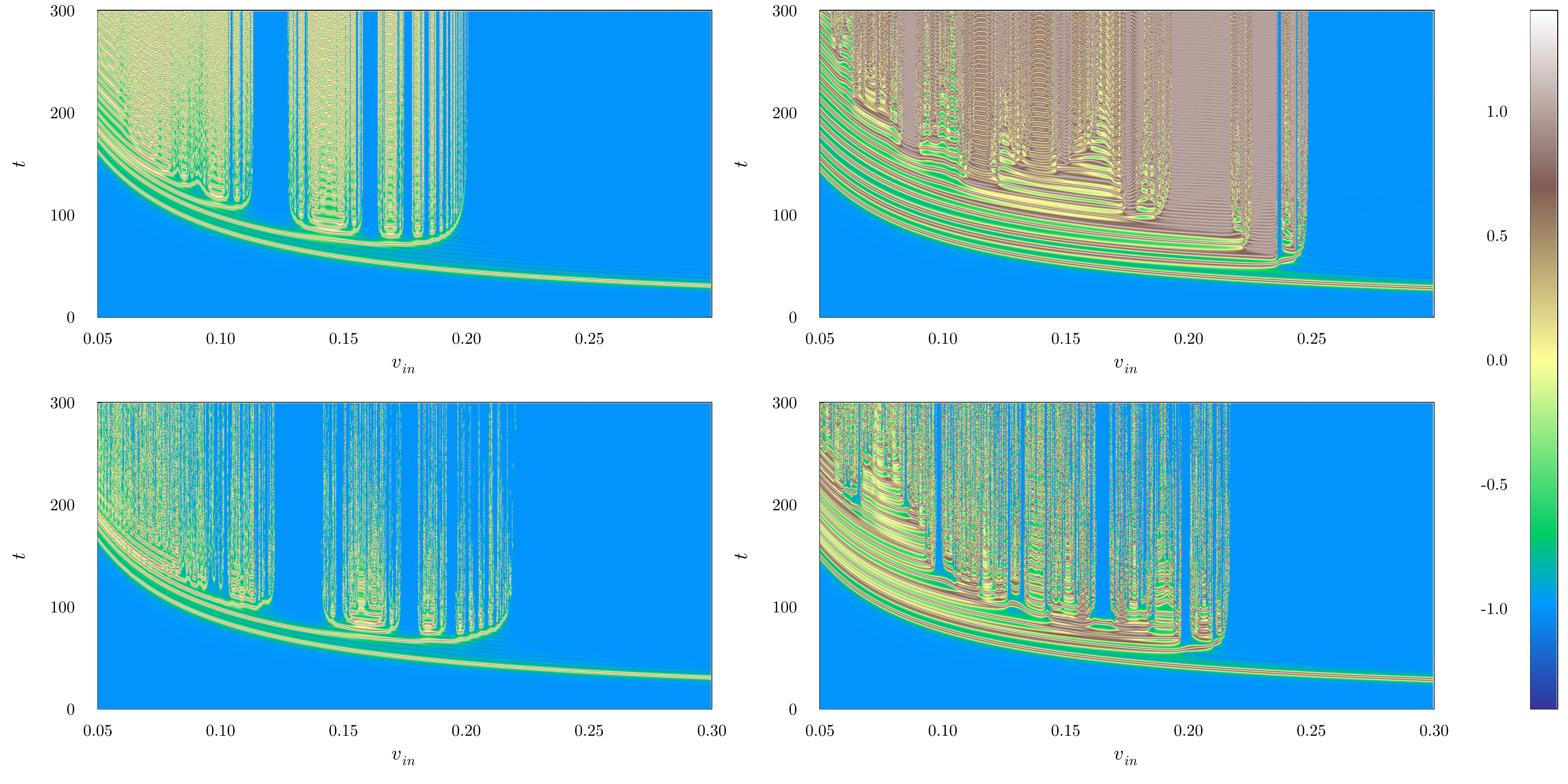}  
   \caption{Kink-antikink collision in the Christ-Lee model. {\it Upper:} full field theory computation for $\epsilon=3$ (left) and $\epsilon=0.5$ (right). {\it Lower:} the CCM result based on pRMS with the first Derrick mode. } \label{KAK-CL-plot}
 \end{figure*}
\begin{figure*}
 \includegraphics[width=1.00\columnwidth]{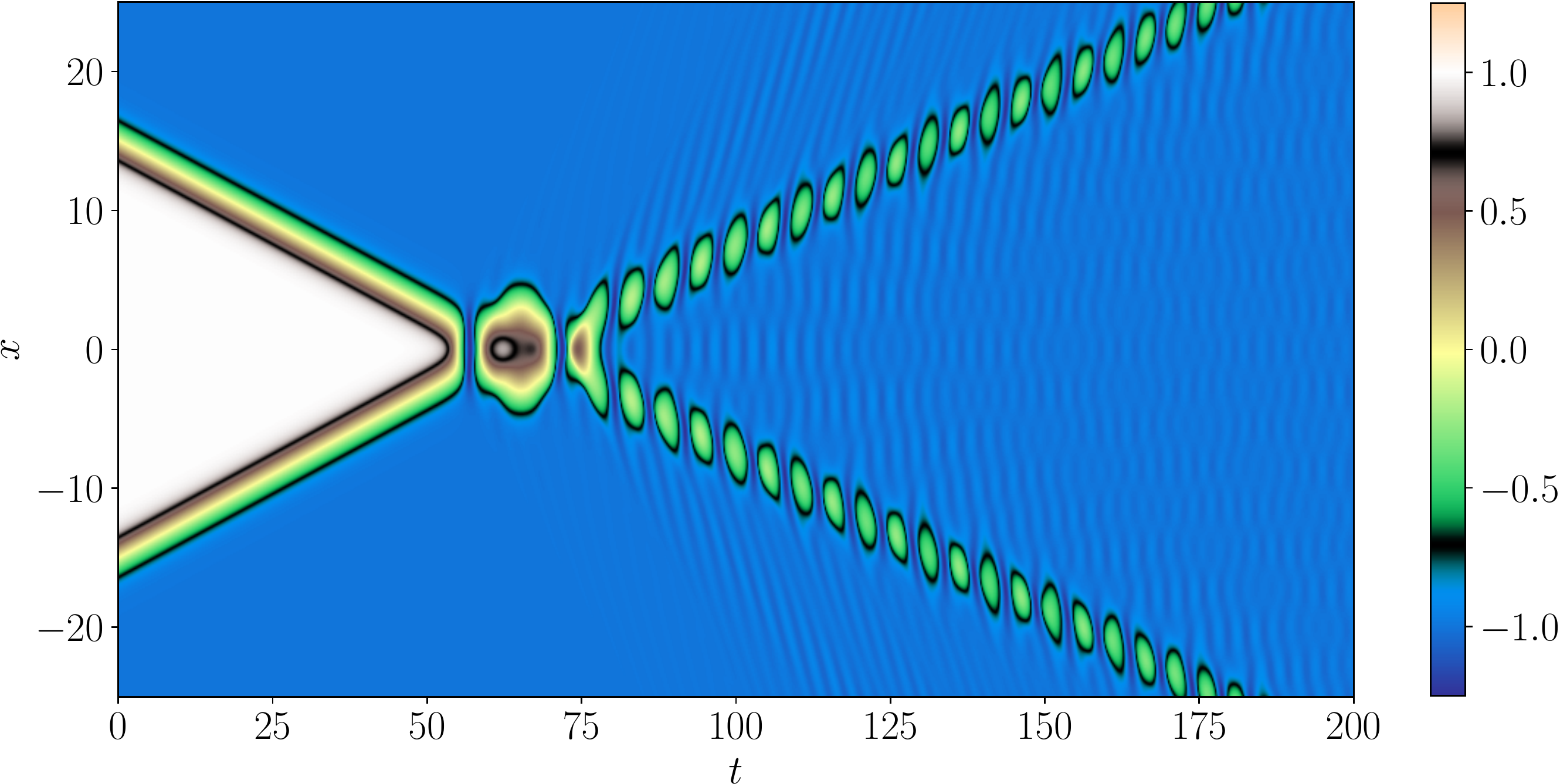} 
     \includegraphics[width=1.00\columnwidth]{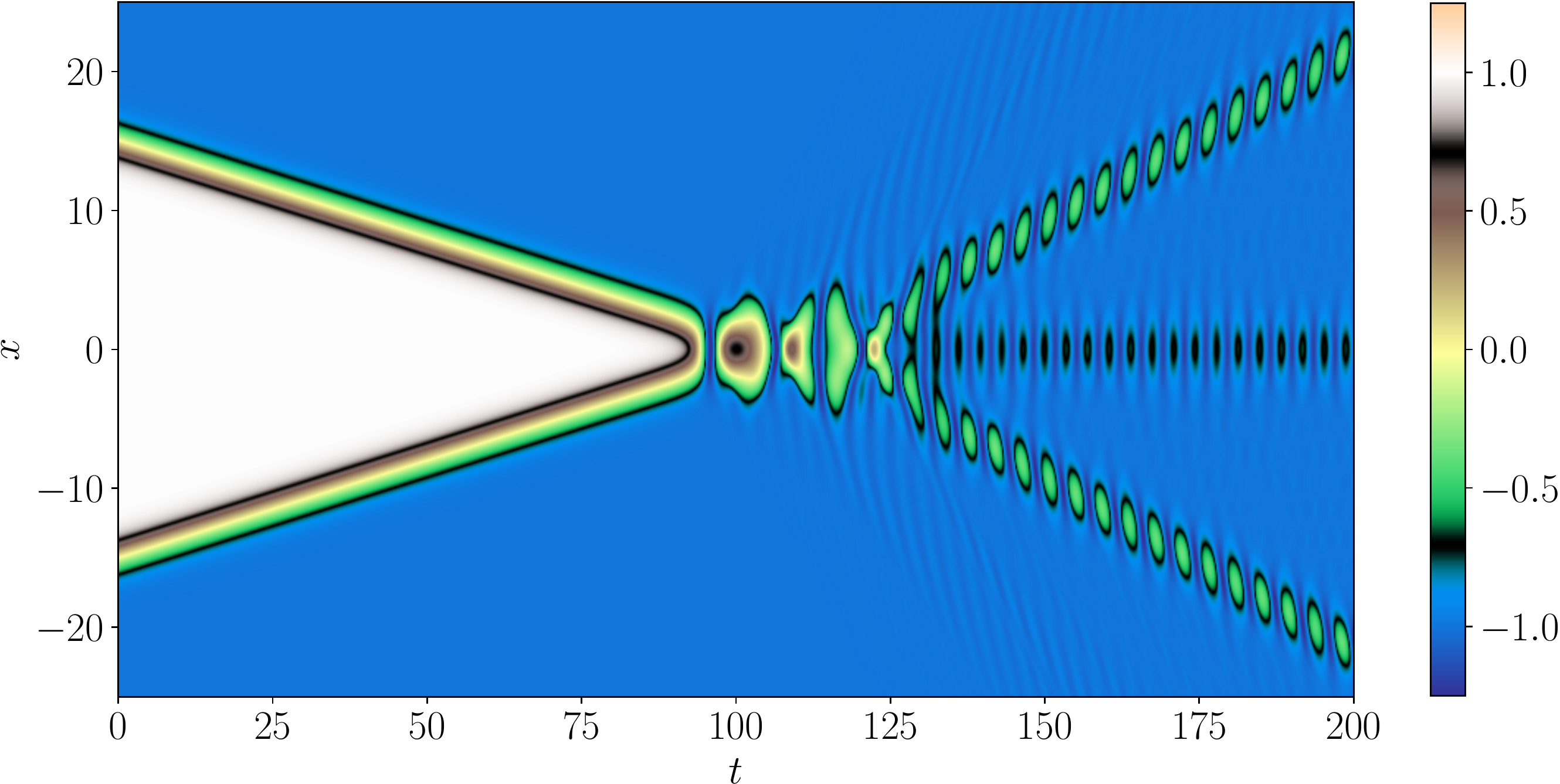} 
   \caption{KAK collision in the Christ-Lee model. Left: annihilation to two oscillons ($\epsilon=0.55, v_{in}=0.24$). Right: annihilation to three oscillons ($\epsilon=0.80, v_{in}=0.14$). } \label{CL-osc}
 \end{figure*}

Here, $\epsilon \in [0, \infty)$ is a parameter which allows for the interpolation between the standard $\phi^6$ model ($\epsilon \to 0$) and the $\phi^4$ model ($\epsilon \to \infty$). For any non-zero $\epsilon$, this theory has two vacua at $\phi=\pm 1$ and, therefore, supports a kink 
\be
\Phi_K=\frac{\epsilon \sinh(x)}{\sqrt{1+\epsilon^2 \cosh^2(x)} }
\ee
and a symmetric antikink $\Phi_{\bar{K}}(x)=- \Phi_K(x)$. As $\epsilon$ decreases to 0, the (anti)kink reveals a composite structure with two well visible centers, which can be interpreted as half-kinks separated by a plateau of the false vacuum with $\phi \approx 0$. The distance between the half-kinks increases as 
$\epsilon \to 0$. In this limit, we obtain two infinitely separated $\phi^6$ kinks, or speaking precisely, the mirror kink $(-1,0)$ and the kink $(0,1)$. Note that the half-kink and half-antikink are not symmetric in our sense, exactly as for the $\phi^6$ kink and antikink. Of course, for any finite $\epsilon$, the half-kinks are confined and cannot be arbitrarily separated to form free states.

Due to the emerging half-kink composite structure arising for small $\epsilon$, kink-antikink collisions in this regime may potentially reveal some additional complexity if compared with the usual multi-kink scattering in $\phi^4$ or $\phi^6$ theories. Indeed, such a collision looks rather like a process of four half-kinks, where each half-kink collision not only excites bound modes but also leads to the emission of radiation, which can affect subsequent collisions. As is well known, collisions between excited solitons are more complicated than in the standard unexcited case \cite{A, Izq-2}. Therefore, it is a very nontrivial task to model them within any CCM. 
 
Another factor which may increase the complexity of the collisions is the mode structure. For $\epsilon=\infty$, where we recover the $\phi^4$ model, there is only one bound mode, the shape mode. This modes exists for all finite $\epsilon$, but its frequency decreases as $\epsilon \to 0$. However, already for arbitrarily large but finite $\epsilon$ another mode shows up from the mass threshold. When $\epsilon$ decreases, more and more bound modes show up. They are of the same nature as the delocalized or trapped two-soliton bound modes observed in antikink-kink collisions in $\phi^6$ theory,  see \cite{Tr} and \cite{phi6}. In fact, the half-kinks are connected by the $\phi=0$ false vacuum, whose mesonic excitations have a smaller mass than the small waves excited in the true vacua. Thus, an effective two-(half)soliton potential well appears. Its width grows with the distance between the half-kinks and, therefore, the number of hosted modes grows without limit as $\epsilon$ decreases to 0.  In Fig. \ref{CL-mode}  we plot the lowest energy mode only. Its frequency changes from $\omega^2=3$ for $\epsilon \to \infty$, where we find just the $\phi^4$ shape mode, to $\omega=0$ for $\epsilon= 0$, where we find two $\phi^6$ kinks, which, if treated separately, do not possess any bound modes. In this limit, this mode tends to another zero mode as two half-kinks become independent kinks at $\epsilon=0$. This behaviour is quite well approximated by the first Derrick mode. As we see in Fig. \ref{CL-mode}, its frequency very well agrees with the frequency of the shape mode until small $\epsilon$ where a growing discrepancy is visible. 

Following these observations, one can expect that KAK scatterings in the Christ-Lee model strongly depend on the value of $\epsilon$, see \cite{CL-Kev}, \cite{Tr-2}. Again, an important observable is the critical velocity which separates one-bounce collisions, where the kink and antikink back scatter to infinities, from more involved dynamics. Obviously, the critical velocity also varies with the parameter of the model. In fact, the observed relation is very nontrivial, as is shown in Fig. \ref{CL-vcrit}, green curve. In the limit $\epsilon \to \infty$ it approaches the $\phi^4$ model value, $v_{cr}=0.2598$. Then, for decreasing $\epsilon$, $v_{cr}$ also decreases until $\epsilon=1.4$ where the critical velocity takes the minimal value, $v_{cr}=0.07$. For even smaller $\epsilon$, $v_{cr}$ begins to grow, reaching a local maximum, and then, again decreases. Such a non-monotonous behaviour repeats as $\epsilon$ tends to 0. 

Now, we apply the pRMS construction to model KAK collisions. In particular, we will focus on the critical velocity. We find that the highly nontrivial $v_{cr}(\epsilon)$ relation is already reproduced quite well by a CCM based on the pRMS with just one Derrick mode. See Fig. \ref{CL-vcrit}, where the blue curve representing the CCM result rises and decreases in good agreement with the PDE computations.  An especially nontrivial fact is that we can quite well approximate the critical velocity even in the regime of a relatively small $\epsilon$, where the kinks begin to exhibit a well pronounced double half-kink structure.  

In general, the inclusion of the second Derrick mode improves the agreement between the CCM and PDE computations, see the orange curve in Fig. \ref{CL-vcrit}, which now lies much closer to the line obtained in the full theory. Thus, once again we find an evidence for the convergence of the presented framework. This is the case for models with $\epsilon >0.4$. For smaller $\epsilon$, the results get worse. This can probably be explained by the fact that in this regime the half-kink structure dominates in the scatterings. Therefore, the addition of the next Derrick mode of the full kink, which treats the two half-kinks forming a kink as one rigid structure, is not related to any actual situation in the full theory. As we already mentioned, in the small $\epsilon$ limit the kink-antikink scattering is, in reality, a four soliton collision where some of the final states can be explained only in terms of half-kink processes. These are, e.g., annihilations of the kink-antikink pair to two and three oscillons, Fig. \ref{CL-osc}. 

We underline that our results are obtained within our general pRMS scheme without the necessity to include any parameter-dependent fitted function \cite{CL-Kev}. The fact that we reproduced the critical velocity curve quite well using only the lowest Derrick normal mode suggests that the higher frequency modes, which appear as $\epsilon$ decreases, may not be the main factors in scattering processes of the Christ-Lee kinks. It seems that, rather, the inner structure of the kinks plays the most significant role \cite{Tr-2}. 

In Fig. \ref{KAK-CL-plot} we also show scans of the KAK collisions for two values of $\epsilon$. It is clearly seen that there is a good qualitative agreement in the single kink regime ($\epsilon=3.0$), where the results resemble the KAK scattering in $\phi^4$ model, as well as in the half-kink regime ($\epsilon=0.5$). Of course, due to the lack of radiation, the CCM has a tendency to provide a larger number of bounce windows instead of the expected bion chimneys. We saw this feature already in the case of the $\phi^4$ model.

\section{Double sine-Gordon model}
 \begin{figure}
 \includegraphics[width=1.00\columnwidth]{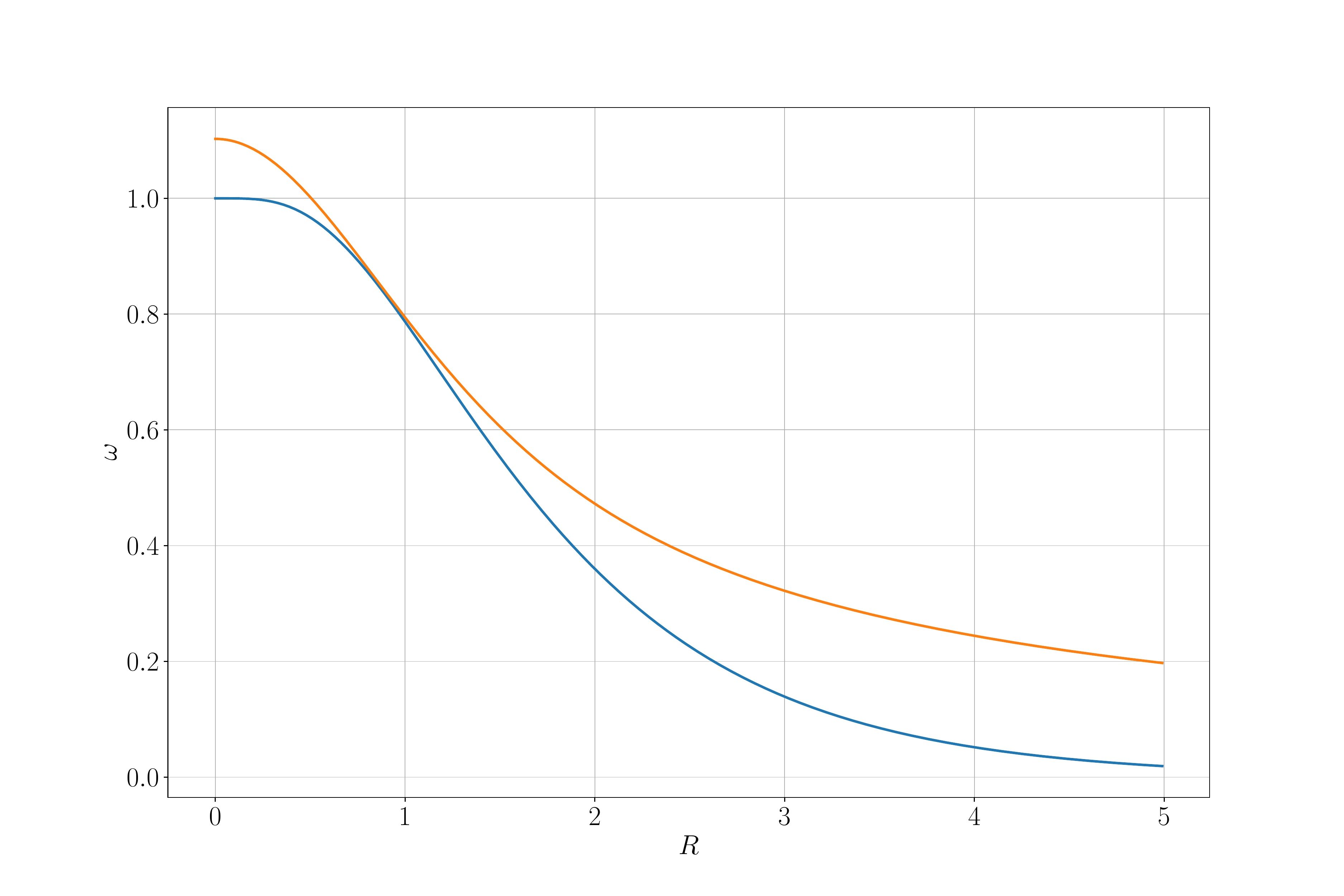} 
   \caption{The bound mode in the double sine-Gordon model (blue) and the first Derrick mode (orange).} \label{2sG-mode}
    \end{figure}
 \begin{figure}
     \includegraphics[width=1.00\columnwidth]{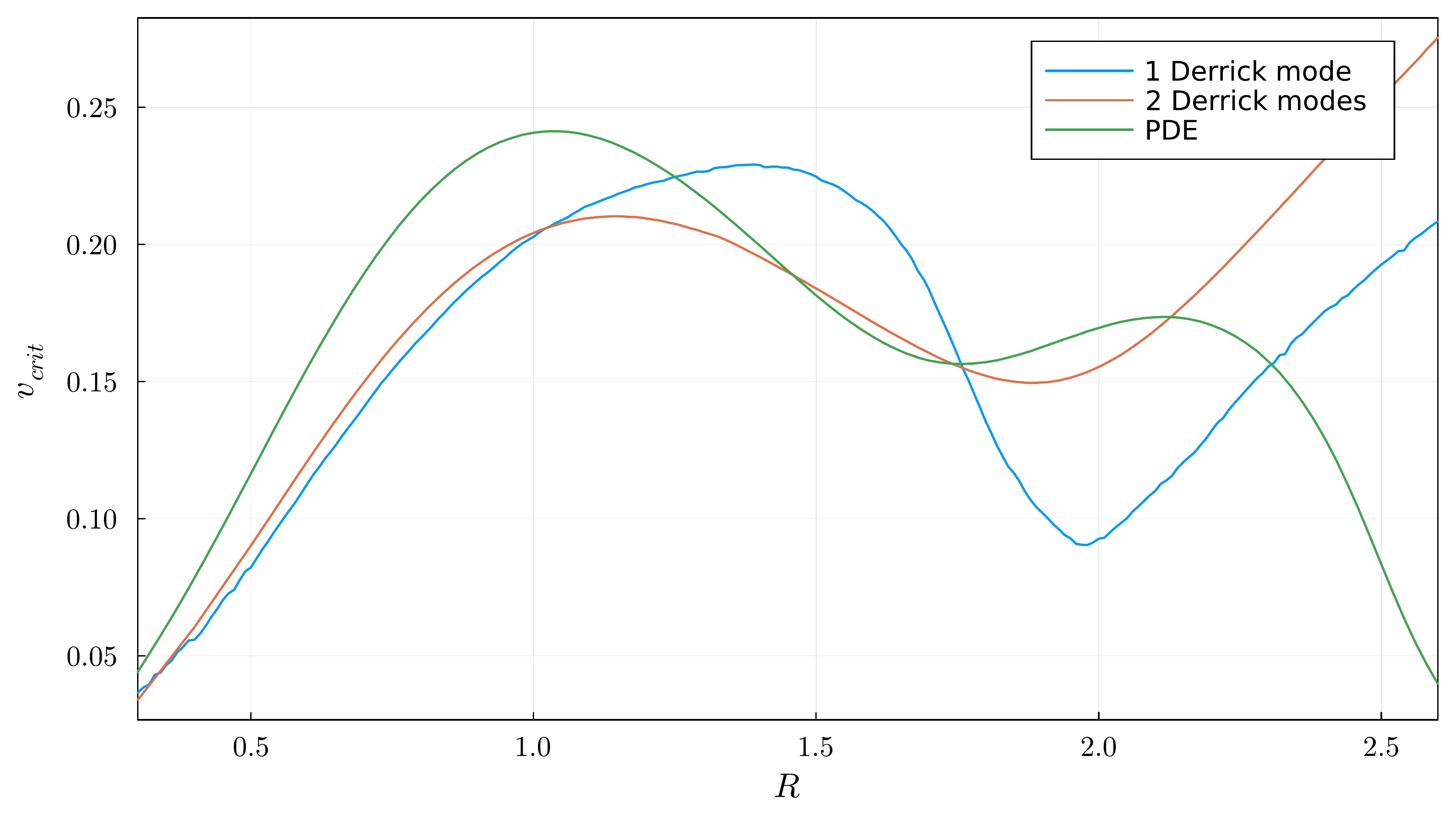} 
   \caption{The critical velocity $v_{cr}$ in the double sine-Gordon model (green) and result obtained from the CCM based on one Derrick mode (blue) and two Derrick modes (orange). } \label{2sG-vcrit}
 \end{figure}
The last family of models we want to analyze is the well-known double sine-Gordon model,
\be
U_{2sG}=\tanh^2R \; (1-\cos \phi) +\frac{4}{\cosh^2 R} \left(1+\cos \frac{\phi}{2} \right),
\ee
where the parameter $R\in [0,\infty)$. This model interpolates between the ordinary sine Gordon, $R\to \infty$, and another (rescaled) sine-Gordon model. The corresponding kink 
\be
\Phi_K = 4\arctan \left( \frac{\sinh x}{\cosh R} \right)
\ee
is in fact a superposition of two sine-Gordon kinks located at $x=\pm R$
\be
\Phi_K = 4\arctan e^{x+R} - 4\arctan e^{R-x}.
\ee
Thus, similarly as in the Christ-Lee model, the kink experiences the appearance of an inner, composite structure. Here, it occurs for growing $R$. 
 \begin{figure*}
  \includegraphics[width=2.0\columnwidth]{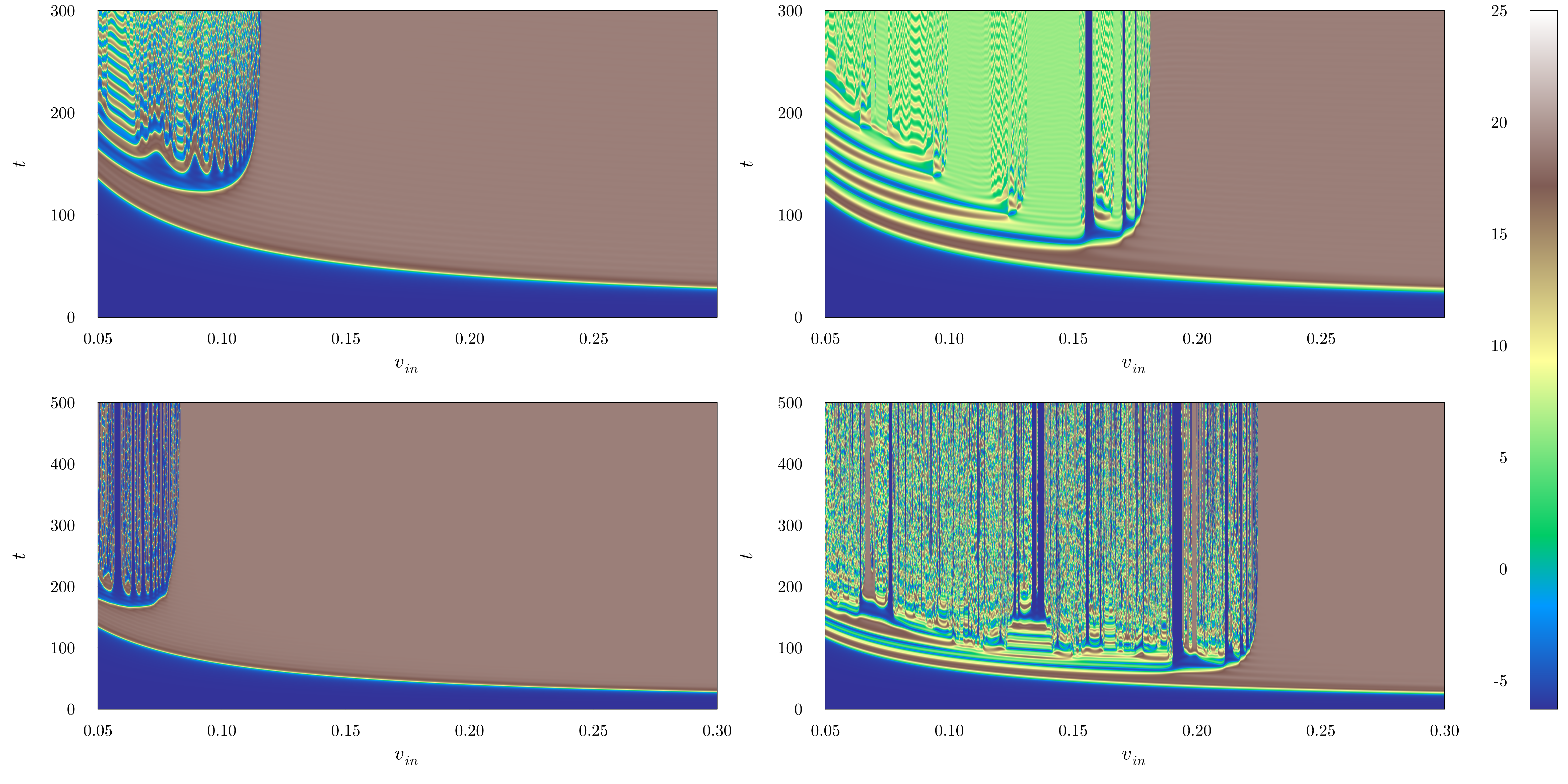} 
   \caption{Kink-antikink collision in the double sine-Gordon model. {\it Upper:} full field theory computation for $R=0.5$ (left) and $R=1.5$ (right). {\it Lower:} the CCM result based on pRMS with the first Derrick model. } \label{KAK-2sG-plot}
 \end{figure*}

If compared with the Christ-Lee model, the main difference is the fact that the double sine-Gordon kink may host only one bound massive mode for all $R\in (0,\infty)$, see Fig. \ref{2sG-mode}, blue curve. Since in the limited cases we obtain the sine-Gordon models, this shape mode disappears as $R\to 0$ and $R\to \infty$. Specifically, for $R\to 0$, the shape mode becomes a non-normalizable threshold mode whose frequency tends to 1, which is the value at which the continuum spectrum begins. For $R\to \infty$, the shape modes approaches the second zero mode since we tend to a model which is a double copy of the sine-Gordon model. 

The bound mode is reasonably well approximated by the Derrick mode, see Fig. \ref{2sG-mode}, orange curve. This works especially well for $R\approx 1$. For a too small $R$, the Derrick mode has its frequency bigger than 1. For $R>2$, where the shape mode can be treated as a quasi-zero mode, the frequency of the Derrick mode is significantly above 0.  This behaviour corresponds to the $\epsilon \to 0$ limit in the Christ-Lee model.

\vspace*{0.2cm}

Kink scattering processes in the double sine-Gordon model have been extensively studied in the literature \cite{CPS, Gani, Gani-2, Sim, Moh}. The main message is that the chaotic structure in the final state formation, which reveals multi-bounce windows immersed between bion chimneys, ceases to exist if $R$ is too small or too large, i.e., if we are too close to the ordinary sine-Gordon model. Concretely, bounce windows are observed for $R \gtrsim 0.5$ and $R \lesssim 2 $. For smaller or bigger values of $R$ the critical velocity rapidly decreases and no chaotic structures are found. Once again, the critical velocity shows a very nontrivial dependence on the parameter of the model, $R$, see Fig. \ref{2sG-vcrit}. Note that here the elastic scattering is a process where the solitons pass through each other and reappear as free final states.

The pRSM once again leads to CCMs which quite well capture the critical velocity. As expected from the previous analysis, this especially concern the regime where the kink is not divided into two half-kink. Thus, when $R \lesssim 2$, see Fig. \ref{2sG-vcrit}, where we compared the $v_{cr}(R)$ curve obtained in the full field theory and in the CCM based on one and two Derrick modes. Even the first Derrick mode qualitatively captures the dependence of $v_{cr}$ on $R$. However, the inclusion of the second Derrick mode significantly improves the agreement. This provides further evidence for the convergence of our perturbative scheme. 

In Fig. \ref{KAK-2sG-plot} we compare the KAK dynamics of the double sine-Gordon model with the CCM based on the first Derrick mode for $R=0.5$ and  $R=1.5$.

\section{Summary}

In the current work, we show that the perturbative Relativistic Moduli Space (pRMS) framework, based on collective coordinates provided by scaling Derrick modes, can be very successfully applied to model kink-antikink collisions in (1+1)-dimensional field theories. In particular, we found that the corresponding CCMs can reproduce the critical velocity $v_{cr}$ amazingly well. $v_{cr}$ is one of the most important quantities characterizing multi-kink collisions, which separates the one bounce regime (where solitons meet only once and then are back scattered or pass through each other) from the multi-bounce regime (where kinks meet several times and, eventually, reappear in the final state as free particles of annihilate to the vacuum). 

Importantly, increasing the set of collective coordinates by the inclusion of a bigger number of Derrick modes, we see that the critical velocity derived in the CCM shows a tendency to converge to the value obtained in the original field theory. Thus, the RMS approach seems to provide a convergent approximation for the critical velocity. To the best of our knowledge this is the {\em first example} of a {\em convergent} perturbative expansion based on the collective coordinates. 

This result applies to theories with two qualitative features:
\begin{itemize}
\item[ {\it i)} ] the solitons do not reveal too well pronounced inner structure, manifested, e.g., as the existence of half-kink substructures;
\item[ {\it ii)} ] the solitons host a well pronounced (not necessary unique) bound mode. 
\end{itemize}  

As the first example, we considered the $\phi^4$ theory where the critical velocity obtained in the CCM with the first four Derrick modes agrees with a more than $99\%$ percent precision with the full field theory result. Next, we computed the critical velocity in the RMS framework for the Christ-Lee and double sine-Gordon family of models and found that the CCM computations capture the non-trivial dependence of $v_{cr}$ on the parameter of the model already if only one Derrick mode is included. The addition of the second mode increases the agreement as long as the kinks do not exhibit a too well visible inner half-kink structure, that is, $\epsilon \gtrsim 0.4$ for the Christ-Lee model and $R \lesssim 2$ for the double sine-Gordon model. 

It should be stressed that, within this range of parameters, the critical velocity is still given by a rather nontrivial, non-monotonous curve, which is very well reproduced in the CCM approach. In fact, at the boundaries of this range one can even recognize the appearance of the inner half-kink structure. Thus, the fact that our approach still keeps its predictive power is even more striking. 

Moreover, our result may indicate that the non-monotonous dependence of the critical velocity on the parameter $\epsilon$ in the Christ-Lee model is probably  related to the appearance of the inner half-kink structure rather than to the growing number of normal modes. This is based on two observations. Firstly, a similar pattern occurs in the double sine-Gordon model where the number of internal modes remains constant for all values of the parameter $R$ which also controls the half-kink structure. Secondly, our CCM model captures this non-monotonous dependence even if it is based only on the first Derrick mode. One can, therefore, conjecture that, in general, for a given field theory, the appearance of an inner substructure in the solitonic solution has probably a bigger impact on the kink dynamics than the existence of a larger number of normal modes.

It is also straightforward to see how one may further improve the CCM description in the composite kink limit. Namely, instead of treating the kink as one particle with its Derrick modes we should rather introduce collective coordinates associated with each half-kink independently. This approach should also apply to composite kinks in other field theories, e.g., \cite{inner-1, inner-2}. We remark that such a construction may be still applicable in the case where the half-kinks are (almost) on top of each other. Indeed, the motion of the half-kink may mimic the shape mode. Interestingly, such a decomposition of a kink into smaller structures resembles the ideas recently proposed in the so-called mechanization program \cite{FB}. 

\hspace*{0.2cm}

There are several interesting directions in which our work may be developed. First of all, one should include more Derrick modes and test the convergence of the critical velocity at higher order. Secondly, one should improve the approach in the half-kink regime along the lines described above. Thirdly, one should use the CCMs obtained here and explain, hopefully in an analytical way, the shape of the $v_{cr}$ curve. Indeed, it would be great to know which features of a kink field theory decide the value of the critical velocity. 

Given their relevance for the understanding of the classical soliton dynamics, one may ask about the role of Derrick modes in the quantum kink models. A natural path for this study can be the recently developed manifestly finite approach by Evslin \cite{JE-1} in which various shape mode--meson processes have been already analyzed \cite{JE-2,JE-3}. Of course, a more direct approach i.e., canonical quantization of the analyzed here CCMs is also an interesting option. It would be very interesting to compare it with recent semicalssical results, see \cite{V}.

\hspace*{0.2cm}

Looking from a wider perspective, our findings provide further solid arguments for the importance of the resonant energy transfer mechanism in solitonic collisions. They also contribute to the recent spectacular improvements in the application of the collective coordinate approach to kink collisions. In fact, quite recently not a single CCM was known which could genuinely model KAK scattering processes, see e.g., \cite{Kev}. Nowadays, the number of theories where CCMs give good or even excellent predictions grows rapidly, see \cite{MORW, RMS, phi6, inst}. All that shows that the moduli space description is a powerful technique, even for the analysis of such extremely complicated processes as kink collisions.

\section*{Acknowledgements}

The 
authors acknowledge financial support from the Ministry of Education, Culture, and Sports, Spain (Grant No. PID2020-119632GB-I00), the Xunta de Galicia (Grant No. INCITE09.296.035PR and Centro singular de investigación de Galicia accreditation 2019-2022), the Spanish Consolider-Ingenio 2010 Programme CPAN (CSD2007-00042), and the European Union ERDF. DC and KO was supported by the Polish National Science Centre 
(Grant NCN 2021/43/D/ST2/01122). 


\appendix
\section{Numerical approach}
Integrating the equations of motion of a CCM is a challenging task from a numerical point of view. Even if both the moduli space metric and the effective potential are known in an analytical form, 
the floating point errors can be the source of many problems \cite{MORW}. 
In the case of the $\phi^4$ model, e.g., subtracting higher order derivatives led to a so-called catastrophic cancellation problem, especially near $a=0$. 
It appeared to be more stable to evaluate the metric and potential numerically \cite{RMS}. Such a procedure reduced the number of numerical artefacts and, surprisingly, did not lead to large computational time overhead. Unfortunately, some numerical problems remained, especially when the Derrick modes are divided by $\tanh(a)$, which is required to resolve the null vector problem. 

One remedy was to expand the profile functions as a Taylor series for $|a|<a_{cut}\ll1$. This, however, complicates the procedure and, what is more important, introduces small discontinuities at $a=\pm a_{cut}$ which can lead to more numerical artefacts (due to the violation of conjectures for existence and uniqueness theorems). 

For the cases where the moduli space has not too large dimension, the best approach is to store the equations of motions in the form of cubic splines. This proved to be very effective even for very complicated profiles such as instantons \cite{inst}. Unfortunately, for moduli spaces with more than two coordinates, such an approach would require storing a large amount of data and the direct integration of equations of motion at each time step was more efficient. Furthermore, the discontinuity issue still remains. 

In order to avoid discontinuities at $a=\pm a_{cut}$ we adopted a different approach. 
Note that the divisor $\tanh(a)$ in (\ref{pRMS-set}) is somewhat arbitrary. Indeed, it can be any smooth function of $a$ which obeys two properties. Namely, it tends to $\pm 1$ as the kink-antkink distance goes to infinity, $\tanh(a) \to \pm 1$ as $a\to \pm \infty$, and it has a linear zero at $a \to 0$. 
This can be achieved, e.g., by replacing the Derrick modes
\bea \label{Derrics}
    && D_k(x, a)=\\ 
    &&\frac{1}{k!} \left( (x+a)^k \Phi_K^{(k)} (x+a) - (x-a)^k \Phi_K^{(k)} (x-a) \right) \nonumber
\eea
with the following smooth approximation
\be
 D_k(x, a) \to \frac{D_k(x)}{\cosh(\alpha a)} + D_k(x, a)\tanh(\alpha a), \label{approxDerrics}
\ee
where $D_k(x)$ is the limit 
\bea
     D_k(x) &=& \lim_{a\to 0}\frac{D_k(x, a)}{a} \nonumber \\
     &=&-\frac{2x^{k-1}}{k!}\left(k\Phi^{(k)}(x)+x\Phi^{(k+1)}(x)\right). 
\eea
We fit the scaling constant $\alpha$ in such a way that (\ref{approxDerrics}) approximates best (\ref{Derrics}). Such an approximation is always regular and smooth and avoids the division by $\tanh(a)$ near $a=0$. This procedure is more robust and almost completely eradicates numerical artefacts.

\end{document}